\documentclass[onecolumn]{elsart3p}
\usepackage{graphicx,amsmath,amssymb,bm,mathptmx,hyperref}

\journal{Mathematics and Computers in Simulation}

\DeclareMathOperator{\tr}{tr}
\DeclareMathOperator{\cn}{cn}
\DeclareMathOperator{\sgn}{sgn}
\DeclareMathOperator{\sech}{sech}

\begin{document}
\begin{frontmatter}

\title{Internal solitary waves in the ocean: Analysis using the periodic, 
inverse scattering transform}

\author{Ivan Christov\thanksref{nu}}
\ead{\href{mailto:christov@alum.mit.edu}{christov@alum.mit.edu}}
\ead[url]{\url{http://alum.mit.edu/www/christov/}}
\address{Code 7181, Naval Research Laboratory, Stennis Space Center, MS 39529-5004, USA}
\thanks[nu]{Present address: Department of Engineering Sciences and Applied Mathematics, Northwestern University, Evanston, IL 60208-3125, USA.}

\begin{abstract}
The periodic, inverse scattering transform (PIST) is a powerful analytical tool in the theory of integrable, nonlinear evolution equations. Osborne pioneered the use of the PIST in the analysis of data form inherently nonlinear physical processes. In particular, Osborne's so-called nonlinear Fourier analysis has been successfully used in the study of waves whose dynamics are (to a good approximation) governed by the Korteweg--de Vries equation. In this paper, the mathematical details and a new application of the PIST are discussed. The numerical aspects of and difficulties in obtaining the nonlinear Fourier (i.e., PIST) spectrum of a physical data set are also addressed. In particular, an improved bracketing of the ``spectral eigenvalues'' (i.e., the $\pm 1$ crossings of the Floquet discriminant) and a new root-finding algorithm for computing the latter are proposed. Finally, it is shown how the PIST can be used to gain insightful information about the phenomenon of soliton-induced acoustic resonances, by computing the nonlinear Fourier spectrum of a data set from a simulation of internal solitary wave generation and propagation in the Yellow Sea.
\end{abstract}


\end{frontmatter}
\section{Introduction}
\label{sec:intro}

The Korteweg--de Vries (KdV) equation is arguably the most famous ``soliton-bearing'' equation, governing phenomena as seemingly disparate as the motion of lattices, the collective behavior of plasmas and the evolution of hydrodynamic waves (see, e.g., \cite{as81,a03,ob80,op94,sep03,zk65,zh99} and the references therein). From a mathematical point of view, the most interesting property of the KdV equation, discovered by the celebrated numerical experiment of Zabusky and Kruskal (ZK)  \cite{zk65}, is the ability of its (localized) traveling-wave solutions to retain their ``identity'' (shape, speed) after colliding with each other. This has been a topic subject to much research over the last few decades,  giving birth to the idea of integrable equations and leading to the discovery of the \emph{inverse scattering transform}. In particular, it has been shown that the KdV equation is ``fully-integrable'' on both the real line and periodic intervals;  therefore, it can be solved \emph{exactly} by the (periodic) inverse scattering transform \cite{as81}. What is more, Salupere et al.~\cite{sep03} have shown that the ZK experiment features more than just the particle-like interactions of the localized solutions of the KdV equation: there exist hidden solitons, and soliton ensembles emerge on long time scales in the ``wave soup'' created by the ZK sine wave initial condition. 

It is this richness of the solution of the \emph{periodic} KdV equation and its integrability by the scattering transform that form the basis of Osborne's \emph{nonlinear Fourier analysis} \cite{oo91}. The latter has been successfully employed in the Fourier-like decomposition of data from inherently nonlinear physical phenomena such as shallow-water ocean surface waves \cite{osbc91}, laboratory-generated surface waves \cite{op94} and internal gravity waves in a stratified fluid \cite{zh99}. In the present work, we focus on the implementation and application of Osborne's nonlinear Fourier analysis to yet another nonlinear-wave phenomenon governed by the KdV equation: internal solitary waves in the ocean.

In their 1980 article, Osborne and Burch \cite{ob80} put forth a model of internal solitary waves in the ocean based on the KdV equation. Since then, their model has been applied to a variety of internal solitary wave phenomena (see, e.g., \cite{a03} and the references therein). In particular, the  latter model was employed by Zhou et al.~\cite{zzr91} to explain the anomalous acoustic signal loss in the presence of internal solitary waves in the ocean. More recently, Chin-Bing et al.~\cite{cb03} and Warn-Varnas et al.~\cite{wv03,wv05} have provided evidence of such anomalous signal losses via coupled ocean-acoustic simulations. However, though there is ample evidence of the effects of internal solitary waves on acoustic signals, there remain unanswered  (theoretical) questions regarding, e.g., energy transfer (mode conversions) in the acoustic field (see, e.g., \cite{wv03} and the references therein). Thus, it is a goal of the present work to lay the foundation for the use of the nonlinear Fourier analysis of internal solitary wave trains in the ocean in obtaining more accurate and ``natural'' wavenumber (or mode) ranges than the ordinary Fourier transform. These, in turn, can be used to prove or disprove conjectures regarding energy transfer in the acoustic field.

This paper is organized as follows. In Section~\ref{sec:intsols_model}, the KdV-based mathematical model of internal solitary waves in the ocean is described. In Section~\ref{sec:ist}, the scattering transform for the KdV equation is presented and its interpretation as a nonlinear generalization of the Fourier transform is discussed. In Section~\ref{sec:numerics}, the numerical implementation of the PIST is addressed, some of the shortcomings of previous algorithms are pointed out, and improvements are suggested. In Section~\ref{sec:results}, the PIST is applied to the analysis of simulated internal solitary wave data and the results are presented. Finally, in Section~\ref{sec:perspectives}, we discuss the future of the present technique.

\section{A mathematical model of internal solitary waves in the ocean}
\label{sec:intsols_model}

\begin{figure}
\centerline{\includegraphics*{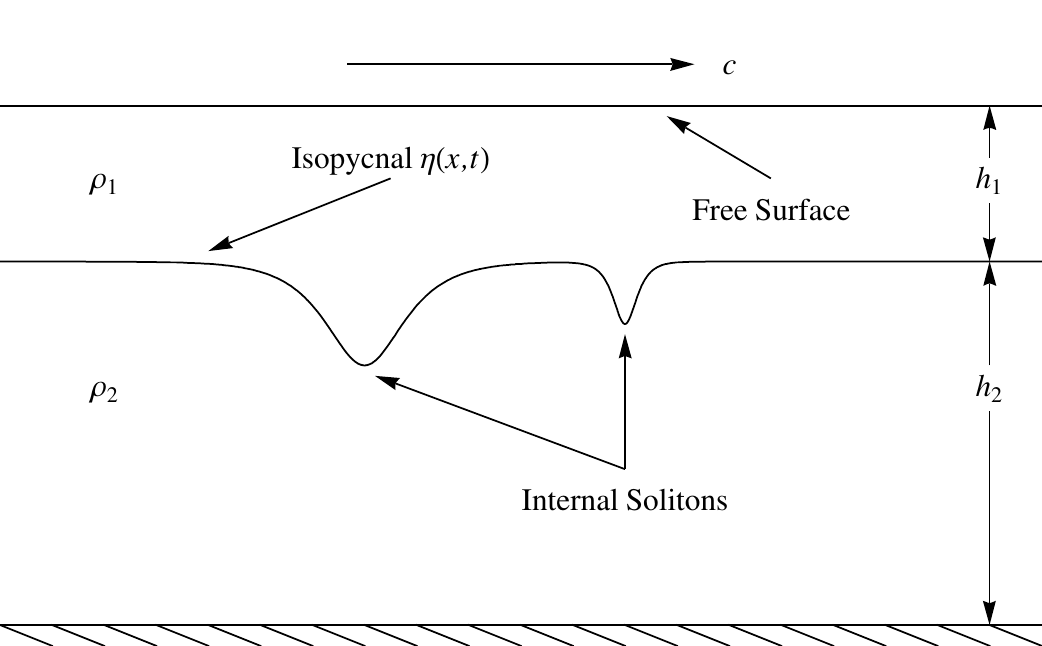}}
\caption{Schematic of the structure of internal solitary waves in the ocean. The notation is as follows: $\rho_1$ and $\rho_2$ denote the fluid densities in the top and bottom layers, respectively; $h_1$ and $h_2$ denote the distances from the \emph{unperturbed} isopycnal to the free surface and to the ocean bottom, respectively; $c$ is a characteristic propagation speed of the solitary-wave packets in the direction indicated by the arrow.}
\label{fig:int_sols}
\end{figure}

In this section, we summarize Osborne and Burch's \cite{ob80} model of internal solitary waves (or solitons, as the case might be) in the ocean. It is based upon the assumption of a two-layer stratification in the ocean, where the two layers are separated either by the \emph{thermocline} or by the \emph{pycnocline}, i.e., a region of rapid change in the temperature or the density of the stratification, respectively. Then, it is supposed that solitary waves traveling on the interface between the two layers are governed by the ubiquitous KdV equation, whose solitary wave solutions \emph{are} solitons (so, the terms `solitary wave' and `soliton' become interchangeable henceforth). The (displaced) interface is a surface of constant temperature or constant (potential) density, and it is thus referred to either as an \emph{isotherm} or as an \emph{isopycnal}, depending on the type of stratification. This physical set up is illustrated in Fig.~\ref{fig:int_sols} (see also Fig.~3 in \cite{ob80}). Moreover, for the remainder of this paper, we restrict ourselves to the case of a desnity-stratified ocean, so we use the terms `pycnocline' and `isopycnal' instead of `thermocline' and `isotherm.'

Under the assumption that the internal solitary waves are ``long'' and the top layer of the stratification is ``shallow'' (this will be made more precise below), the governing (KdV) equation of the isopycnal displacement, which we denote by $\eta(x,t)$, is
\begin{equation}
\eta_t+c_0\eta_x+\alpha\eta\eta_x+\beta\eta_{xxx}=0,\quad (x,t)\in[0,L]\times(0,\infty),
\label{eq:kdv}
\end{equation}
where $L(>0)$ is the length of the domain, and the subscripts denote partial differentiation with respect to an independent variable. In addition, $c_0(>0)$, $\alpha(<0)$ and $\beta(>0)$ are (constant) physical parameters (see, e.g., \cite{a03} for their interpretation). They can be written in terms of the variables in Fig.~\ref{fig:int_sols} as follows \cite{ob80}:
\begin{equation}
c_0 \simeq \sqrt{g\left(\frac{\rho_2-\rho_1}{\rho_1}\right)\left(\frac{h_1h_2}{h_1+h_2}\right)},\quad
\alpha \simeq \frac{-3c_0}{2}\left(\frac{h_2-h_1}{h_1h_2}\right),\quad
\beta \simeq \frac{1}{6}c_0h_1h_2,
\label{eq:kdv_params}
\end{equation}
where $h_1<h_2$ and $\rho_1<\rho_2$. Furthermore, we are interested in the periodic Cauchy problem (i.e., the initial-value problem subject to periodic boundary conditions). Thus, we have that
\begin{equation}
\left\{\begin{aligned}
&\eta(x,0)=\eta_0(x), &&\quad x\in[0,L],\\
&\eta(x+L,t)=\eta(x,t), &&\quad (x,t)\in[0,L]\times[0,\infty).
\end{aligned}\right.
\label{eq:kdv_ic}
\end{equation}
In practice, however, data is rarely periodic, so considering \emph{quasiperiodic} (or, more precisely, ``almost periodic'') solutions, i.e., those  for which $\forall\delta>0$ $\exists \mathfrak{L}=\mathfrak{L}(\delta)$ such that $|\eta(x+\mathfrak{L},t)-\eta(x,t)|<\delta$ for all $(x,t)\in[0,L]\times[0,\infty)$, might be more appropriate. Though the details of the PIST carry over to this case, difficulties emerge in the physical interpretation of the results \cite{o93}. Therefore, we restrict ourselves to periodic data sets and solutions.

Here, we note that more sophisticated \emph{continuous stratification} models (still in conjunction with the KdV equation) have been proposed (see, e.g., \cite{a03} and the references therein) and successfully used in practical modeling \cite{wv03}. However, for the purposes of this work, the well-tested model of Osborne and Burch is sufficient. 

Finally, it is important to outline the limits of applicability of the KdV equation to stratified fluids. The assumption of ``long, shallow-water'' waves we made above can be made more precise by requiring that (see \cite{a03,ob80}) the largest wave amplitude is much smaller than the top layer's depth, i.e., $[\max_{x,t}\eta(x,t)]/h_1 \ll 1$, and that the characteristic width of the waves is much greater than the top layer's depth, i.e., $h_1/W \ll 1$, where $W$ can be taken to be, e.g., the largest half-width of the waves. In practice, as illustrated by the results below, the former condition is far more difficult to satisfy than the latter.

\section{Interpretation of the PIST as a ``nonlinear Fourier transform''}
\label{sec:ist}

Next, we turn to the relationship between the periodic, inverse scattering transform (PIST) and the (ordinary) Fourier transform, and the interpretation of the former as a nonlinear generalization of the latter. To this end, first we note that the solution strategy by the scattering transform can be split into two distinct steps: the \emph{direct problem} and the \emph{inverse problem}. The former consists of solving the eigenvalue problem
\begin{equation}
\mathcal{H}\psi = E\psi,\quad \mathcal{H} := -\frac{\partial^2}{\partial x^2}-\lambda\eta_0(x),
\label{eq:schro}
\end{equation} 
where $\lambda:=\alpha/(6\beta)$ is a nonlinearity-to-dispersion ratio, and $E\in\mathbb{R}$ is a spectral eigenvalue such that $\sqrt{E}\in\mathbb{C}$ is a wavenumber. Quite serendipitously, Eq.~\eqref{eq:schro} has been studied extensively in the literature: in quantum mechanics, $\mathcal{H}$ is the celebrated (time-independent) Schr\"odinger operator, and, in the theory of ODEs, it is known as Hill's operator (a special case of the well-known Sturm--Liouville operator) \cite{c99}. For periodic ``potentials,'' i.e., when $\eta_0(x)=\eta_0(x+L)$ $\forall x\in[0,L]$ as we have assumed, it is well-known that the spectrum of the operator $\mathcal{H}$ is divided into two distinct sets depending upon the boundary conditions imposed on the eigenfunctions $\psi$ \cite{o94,as81}. Thus, it is common to classify the spectral eigenvalues (often referred to as the ``scattering data'') as belonging to either the \emph{main spectrum}, which we write as the set $\{\mathcal{E}_j\}_{j=1}^{2N+1}$, or the \emph{auxiliary spectrum}, which we write as the set $\{\mu_j^0\}_{j=1}^{N}$, where $N$ is the number of degrees of freedom (i.e., oscillations modes).

On the other hand, the inverse problem consists of constructing the \emph{nonlinear Fourier series} from the spectrum $\{\mathcal{E}_j\}\cup\{\mu_j^0\}$ using Abelian hyperelliptic functions \cite{os90,o93} or the Riemann $\Theta$-function \cite{bh91,o95}. In former case, which is the so-called $\mu$-representation of the PIST, the \emph{exact} solution of Eq.~\eqref{eq:kdv}, subject to the initial and boundary conditions given in Eq.~\eqref{eq:kdv_ic}, takes the form
\begin{equation}
\eta(x,t)=\frac{1}{\lambda}\left\{2\sum_{j=1}^N \mu_j(x,t) - \sum_{j=1}^{2N+1} \mathcal{E}_j\right\}.
\label{eq:nl_fs}
\end{equation}
It is important to note that all \emph{nonlinear} waves and their \emph{nonlinear} interactions are accounted for in this \emph{linear} superposition. Unfortunately, the computation of the nonlinear oscillation modes (i.e., the hyperelliptic functions $\{\mu_j(x,t)\}_{j=1}^N$) is \emph{highly} nontrivial; however, numerical approaches have been developed \cite{os90} and successfully used in practice \cite{osbc91,op94}. In addition, we note that the auxiliary spectrum, often referred to as the hyperelliptic function ``phases,'' is such that $\mu_j^0\equiv\mu_j(0,0)$ \cite{o94,op94}.

Several special cases of Eq.~\eqref{eq:nl_fs} offer insight into why the latter is analogous to the ordinary Fourier series. In the small-amplitude limit, i.e., when $\max_{x,t}|\mu_j(x,t)|\ll1$, we have $\mu_j(x,t)\sim \cos(k_j x-\omega_j t+\phi_j)$, where $k_j$ is the wavenumber, $\omega_j$ is the frequency and $\phi_j$ the phase of the mode. Therefore, if we suppose that all the oscillation modes fall in the small-amplitude limit, then Eq.~\eqref{eq:nl_fs} reduces to the ordinary Fourier series! This relationship is more than just an analogy, a rigorous derivation of the (ordinary) Fourier transform from the scattering transform, in the small-amplitude limit, is given in \cite{o91}. Next, if there are no interactions, e.g., the spectrum consists of a single wave (i.e., $N=1$), we have $\mu_1(x,t) = \cn^2(k_1 x-\omega_1 t+\phi_1|m_1),$ which is a Jacobian elliptic function with modulus $m_1$. In fact, it is the well-known \emph{cnoidal wave} solution of the (periodic) KdV equation \cite{as81}.

For the hyperelliptic representation of the nonlinear Fourier series, given by Eq.~\eqref{eq:nl_fs}, the wavenumbers are \emph{commensurable} with those of the ordinary Fourier series, i.e., $k_j=2\pi j/L$ ($1\le j\le N$) \cite{o93}. However, this is not the only way to classify the nonlinear oscillations. One can use the modulus $m_j$, termed the ``soliton index,'' of each of the hyperelliptic functions, which can be computed from the discrete spectrum as 
\begin{equation}
m_j=\frac{\mathcal{E}_{2j+1}-\mathcal{E}_{2j}}{\mathcal{E}_{2j+1}-\mathcal{E}_{2j-1}},\quad 1\le j\le N.
\label{eq:modulus}
\end{equation}
Then, each nonlinear oscillation mode can be placed into one of two distinct categories based on its soliton index:
\begin{enumerate}
	\item $m_j\gtrsim0.99$ $\Rightarrow$ solitons, in particular, $\cn^2(x|m=1)=\sech^2(x)$;
	\item $m_j\ll1.0 \Rightarrow$ radiation, in particular, $\cn^2(x|m=0)=\cos^2(x)$.
\end{enumerate}
For $m_j$ not in either distinguished limit above, the qualitative structure of the nonlinear oscillations is not immediately obvious. Boyd \cite{b82} argues that for moderate moduli the polycnoidal wave solutions of the KdV equation are actually well-approximated by the first few terms in their Fourier series, showing they are, in fact, not much different from the linear waves of the $m_j\ll 1.0$ limit.

Furthermore, it can be shown \cite{osbc91,op94} that the amplitudes of the hyperelliptic functions are given by
\begin{equation}
A_j = \left\{\begin{alignedat}{2} &\tfrac{2}{\lambda}(E_{\mathrm{ref}}-\mathcal{E}_{2j}),&&\quad \text{for solitons};\\ &\tfrac{1}{2\lambda}(\mathcal{E}_{2j+1}-\mathcal{E}_{2j}), &&\quad \text{otherwise (radiation)}.\end{alignedat}\right.
\end{equation}
where $E_{\mathrm{ref}}=\mathcal{E}_{2j^*+1}$ is the \emph{soliton reference level} with $j^*$ being the largest $j$ for which $m_j\ge 0.99$. Then, clearly, the number of solitons in the spectrum is $N_{\mathrm{sol}}\equiv j^*$.

\section{Numerical implementation of the PIST} 
\label{sec:numerics}

In the present work, we have implemented a modified version of Osborne's automatic algorithm \cite{o94} for computing the spectral eigenvalues, which we describe in this section. To this end, we begin the solution of the periodic-potential eigenvalue problem, given by Eq.~\eqref{eq:schro}, by introducing an ``appropriate'' basis of eigenfunctions and appealing to Floquet's theorem \cite{c99} to write the eigenfunctions one period ahead as a linear combination of the current ones:
\begin{equation}
\boldsymbol{\Phi}(x+L,{E}) = \boldsymbol{\alpha}(x+L,{E}) \boldsymbol{\Phi}(x,{E})\quad\forall x\in[0,L],
\end{equation}
where $\boldsymbol{\alpha}(x+L,E)$ is the \emph{monodromy matrix}, and $\boldsymbol{\Phi}$ is the \emph{fundamental matrix} of linearly-idependent eigenfunctions. That is to say, $\boldsymbol{\alpha}$ ``advances'' the eigenfunctions one period to the right, and $\boldsymbol{\Phi}=\begin{pmatrix}\phi &\phi_x\\ \phi^* &\phi_x^*\end{pmatrix}$, for some eigenfunction $\phi$, where a star superscript denotes complex conjugation. Then, the spectral eigenvalues can be related to the monodromy matrix (see, e.g., \cite{o93,o94} and the references therein) as follows:
\begin{align}
&\text{Main Spectrum: } \mathcal{E}_j \text{ such that } \tfrac{1}{2}\tr\boldsymbol{\alpha}(x,{\mathcal{E}_j}) = \pm 1,\\
&\text{Auxiliary Spectrum: } \mu_j^0 \text{ such that } \alpha_{21}(x,{\mu_j^0}) = 0.
\end{align}
Now, this may seem like a circular definition because we need to know the eigenfunctions (and eigenvalues) to find the monodromy matrix. However, it is possible to find a numerically-accessible quantity that we can use to determine $\boldsymbol{\alpha}$ without explicitly finding the eigenfunctions. Then, we can compute the spectrum $\{\mathcal{E}_j\}\cup\{\mu_j^0\}$ of our data set, i.e., of the potential $-\lambda\eta_0(x)$.

To this end, we rewrite the Schr\"odinger eigenvalue problem, given by Eq.~\eqref{eq:schro}, as a first-order system:
\begin{equation} 
\frac{\partial}{\partial x} \Psi(x,E) = {\bm B}(x,E)\Psi(x,E),\quad {\bm B}(x,E) = \begin{pmatrix} 0 & 1 \\ -q(x,E) & 0 \end{pmatrix},
\label{eq:schro_sys}
\end{equation}
where $\Psi:=(\psi,\psi_x)^\top$ and $q(x,E) := \lambda\eta_0(x)+E$. Integrating Eq.~\eqref{eq:schro_sys} with respect to $x$ (while keeping $E$ fixed and treating it as a parameter), we have 
\begin{equation}
\Psi(x,E) = \exp\left\{\int_{x_0}^{x}{\bm B(\tilde{x},E)}\,\mathrm{d}\tilde{x}\right\},
\label{eq:psi_int}
\end{equation}
where $x_0\in[0,L]$ is an arbitrary \emph{base point} \cite{o94}; here, we set $x_0=0$ without loss of generality.

Now, $\eta_0(x)$ is usually a discrete data set, i.e., its values are only known for $x=x_i:=i\Delta x$, where $i$ is an integer such that $0\le i\le M-1$, $M$ is the number of points in the data set and $\Delta x := L/M$. Therefore, for such a \emph{piecewise-constant potential} \cite{o91}, we immediately have [from Eq.~\eqref{eq:psi_int}] that 
\begin{equation}
\Psi(x_{i+1},E) = e^{\Delta x {\bm B}(x_i,E)}\Psi(x_i,E)\quad\forall i.
\label{eq:psi_pwc_potent}
\end{equation}
Iterating the latter relation gives $\Psi(x_0+L,E) = {\bm M}(x_0+L,E) \Psi(x_0,E)$, where 
\begin{equation}
{\bm M}(x_0+L,E) := \prod_{i=M-1}^0e^{\Delta x{\bm B}(x_i,E)}
\end{equation}
is the so-called \emph{scattering matrix}. Also, the matrix exponential above can easily be computed:
\begin{equation}
e^{\Delta x{\bm B}(x_i,E)} = \begin{pmatrix} \cos\Bigl(\Delta x\sqrt{q(x_i,E)}\Bigr) & \sin\Bigl(\Delta x\sqrt{q(x_i,E)}\Bigr)\Big/\sqrt{q(x_i,E)} \\[2mm] -\sqrt{q(x_i,E)}\sin\Bigl(\Delta x\sqrt{q(x_i,E)}\Bigr) & \cos\Bigl(\Delta x\sqrt{q(x_i,E)}\Bigr) \end{pmatrix}\quad\forall i.
\end{equation}
Furthermore, notice that $\sqrt{q(x_i,E)}$ is either real or purely imaginary, meaning that the above matrix is always real. This observation allows for the implementation of the entire algorithm using real arithmetic, speeding it up by a factor of four \cite{o94}. Finally, it can be shown that $\tfrac{1}{2}\tr\boldsymbol{\alpha} \equiv \tfrac{1}{2}\tr{\bm M}$ and $\alpha_{21} \equiv M_{21}$ \cite{o94,op94}. Thus, the scattering matrix $\bm{M}$ is a numerically-accessible analogue of the monodromy matrix $\boldsymbol{\alpha}$ that does \emph{not} require a priori knowledge of the eigenfunctions of $\mathcal{H}$. So, we can use $\bm{M}$ to compute the spectral eigenvalues.

Unfortunately, even with this knowledge, computing the spectrum $\{\mathcal{E}_j\}\cup\{\mu_j^0\}$, i.e., finding the $\pm1$ crossings of the \emph{Floquet discriminant} $\Delta(E) := \tfrac{1}{2}\tr{\bm M}(x_0+L,E)$ and the zero crossings of $M_{21}(E)$, is no easy task. Osborne \cite{o94} suggests using an ``accounting function'' such as 
\begin{equation}
S[\xi(E)] = \sum_{i=0}^{M-2} \tfrac{1}{2}\left| \sgn[\xi(x_{i+1},E)] - \sgn[\xi(x_i,E)] \right|,
\end{equation}
where $\sgn(\chi) = \chi/|\chi|$ for all real $\chi\ne0$ and it vanishes otherwise, to count the zeros of some function $\xi(E)$[$=M_{12}(E)$, $\Delta(E)$, etc] at a certain energy level $E$. Unfortunately, the accounting function $S[\,\cdot\,]$ suffers numerical instability for the spectral scales that result from the  internal solitons model (notice the spike in $S[\Delta(E)]$ in Fig.~\ref{fig:ist_fd} for $E\in[-0.00045,-0,0004]$). The latter instability makes the use of such accounting functions undesirable when $\lambda\lll1$. For example, in the study of internal solitary waves in far deeper water columns than considered here, the nonlinearity-to-dispersion ratio is $|\lambda|\simeq 10^{-8}$ \cite{hwvc07}, and $S[\Delta(E)]$ is dominated by numerical noise. 

Therefore, in the present work, we take a more pragmatic approach to computing the spectral eigenvalues. Rather than using the automatic algorithm, which (in theory) assures that all eigenvalues are found, we simply define a fine enough grid a priori, e.g., $\{E_{\mathrm{min}}+k\Delta E\}_{k=0}^{N_E}$ with $\Delta E:=(E_{\mathrm{max}}-E_{\mathrm{min}})/N_E$, where $N_E$ is the number of intervals and $E_{\mathrm{max}}$, $E_{\mathrm{min}}$ are appropriately chosen upper and lower bounds on the spectral eigenvalues (see below), so that all zero crossings of some function $\xi(E)$[$=M_{12}(E)$, $\Delta(E)$, etc] are resolved. Then, by sampling $\xi(E)$ on the latter grid and checking for sign changes, we obtain a bracketing of its roots. 

Now, the automatic algorithm given in \cite{o94} requires using $S[M_{11}(E)]$ to bracket the zeros of $M_{11}(E)$, which along with the zeros of $M_{21}(E)$  give a bracketing of the $\pm 1$ crossings of $\Delta(E)$. However, since the number of zeros of $M_{21}(E)$ is equal to the number of zeros of $M_{11}(E)$, which is, in turn, equal to the number of oscillations modes in the nonlinear Fourier series \cite{o94}, we only compute the zeros of $M_{21}(E)$, thereby reducing the computational cost of the algorithm. Also, at this point, we recall the following important facts stemming from the structure of the equations \cite{as81}:
\begin{enumerate}
	\item $M_{21}(E)=0$ \emph{exactly once} for $E\in[a,b]$, where $a$ and $b$ are such that $\Delta(a)=\Delta(b)=+1$ or $\Delta(a)=\Delta(b)=-1$, 
	\item $\Delta(E)=0$ \emph{exactly once} for $E\in[a,b]$, where $a$ and $b$ are such that $\Delta(a)=+1$ and $\Delta(b)=-1$ or $\Delta(a)=-1$ and $\Delta(b)=+1$. 
\end{enumerate}
This means that once we have computed all the zeros of $M_{21}(E)$ and $\Delta(E)$ in $[E_{\mathrm{min}},E_{\mathrm{max}}]$, and ordered them in ascending order, we have a complete bracketing of the $\pm 1$ crossings of the Floquet discriminant.

Given the highly oscillatory nature of the $\Delta(E)$ and $M_{21}(E)$ (see Fig.~\ref{fig:ist_fd}), keeping the roots bracketed during the computations is \emph{critical} to the success of the algorithm. Originally, Newton's method was used in \cite{op94}, but, due to the latter's tendency to ``jump out'' of the initial bracket, the bisection method was employed in \cite{o94}. The latter guarantees that the new approximation to the root remains in the initial bracket, but converges only linearly, unlike Newton's method. To get the ``best of both worlds,'' in the present work, we employ the (modified or \emph{Illinois-type}) \emph{regula falsi} method \cite{db74}. The latter guarantees that successive approximations of the root remain in the initial bracket and converges superlinearly. Now, let us summarize our implementation of the regula falsi method  for the reader's convenience. 

Consider a bracket $[E_j^{\mathrm{L}},E_j^{\mathrm{R}}]$ containing the $j$th root of some function $\xi(E)$[$=M_{12}(E)$, $\Delta(E)$, etc]. Then, we begin each iteration of the root-finding algorithm by computing
\begin{equation}
E_j^{\mathrm{C}} = \frac{E_j^{\mathrm{L}}\vartheta^{\mathrm{R}}\xi(E_j^{\mathrm{R}})-E_j^{\mathrm{R}}\vartheta^{\mathrm{L}}\xi(E_j^{\mathrm{L}})}{\vartheta^{\mathrm{R}}\xi(E_j^{\mathrm{R}})-\vartheta^{\mathrm{L}}\xi(E_j^{\mathrm{L}})},
\label{eq:new_root}
\end{equation}
where $\vartheta^{\mathrm{L}}$ and $\vartheta^{\mathrm{R}}$ are weights necessary to ensure that the regula falsi method converges superlinearly. Then, we proceed as follows:
\begin{equation}
\begin{aligned}
&\text{If } \sgn[\xi(E_j^{\mathrm{C}})] = \sgn[\xi(E_j^{\mathrm{R}})] \text{, then }
\left\{\begin{aligned}
&\text{If } \sgn[\xi(E_j^{\mathrm{C,old}})]=\sgn[\xi(E_j^{\mathrm{R}})] \text{, then } \vartheta^{\mathrm{L}}=\tfrac{1}{2} \text{, else } \vartheta^{\mathrm{L}}=1;\\
&E_j^{\mathrm{R}}=E_j^{\mathrm{C}},\; \vartheta^{\mathrm{R}}=1.
\end{aligned}\right\},\\
&\text{Else if } \sgn[\xi(E_j^{\mathrm{C}})] = \sgn[\xi(E_j^{\mathrm{L}})], \text{ then }
\left\{\begin{aligned}
&\text{If } \sgn[\xi(E_j^{\mathrm{C,old}})]=\sgn[\xi(E_j^{\mathrm{L}})] \text{, then } \vartheta^{\mathrm{R}}=\tfrac{1}{2} \text{, else } \vartheta^{\mathrm{R}}=1;\\
&E_j^{\mathrm{L}}=E_j^{\mathrm{C}},\; \vartheta^{\mathrm{L}}=1.
\end{aligned}\right\},\\
&\text{Else } \{\text{ Iteration has converged or bracket was lost. }\},
\end{aligned}
\label{eq:iteration}
\end{equation}
where $E_j^{\mathrm{C,old}}$ is the result of Eq.~\eqref{eq:new_root} from the \emph{previous} iteration. After the conditional statements in Eq.~\eqref{eq:iteration} are evaluated, we have a new bracket for the root $E_j$ of $\xi(E)$, and the algorithm continues by generating a new guess via Eq.~\eqref{eq:new_root}. The iteration is continued until the bracket is ``small enough,'' i.e., $|E_j^{\mathrm{L}} - E_j^{\mathrm{R}}|<\epsilon$, where $\epsilon$ is a tolerance, which we took to be $10^{-12}$ when computing the zeros of $\Delta(E)$ and $10^{-14}$ when computing the zeros of $M_{21}(E)$ and the $\pm1$ crossings of $\Delta(E)$.

For the data analysis in the following section, we chose $E_{\mathrm{min}}=-|\lambda|(\eta_{\mathrm{max}}-\bar{\eta})$, where $\eta_{\mathrm{max}}=\max_x |\eta_0(x)|$ and $\bar{\eta}$ is the (arithmetic) mean of $|\eta_0(x)|$, $E_{\mathrm{max}}=0$ and $N_E=1000$. Note that, while $E_{\mathrm{min}}$ is chosen so that \emph{no} eigenvalues exist to the left of it \cite{o94}, the value of $E_{\mathrm{max}}$ is an empirical estimate based on the spectrum of the data set under consideration. Theoretically, $E_{\mathrm{max}}$ can be taken to be the Nyquist cutoff, i.e., $(\pi/\Delta x)^2$, however, the latter is very large and impractical. We found that $E_{\mathrm{max}}=E_{\mathrm{min}}+2|E_{\mathrm{min}}|$ is sufficiently large for most problems, when there is no clear choice for the value of $E_{\mathrm{max}}$. Finally, we note that this modified numerical PIST algorithm has been benchmarked against the standard examples discussed \cite{ob86}, and its results are in exact agreement those of the with previous versions/impelementations.

\section{Results and discussion}
\label{sec:results}

\begin{figure}
\centerline{\includegraphics*[width=0.48\textwidth]{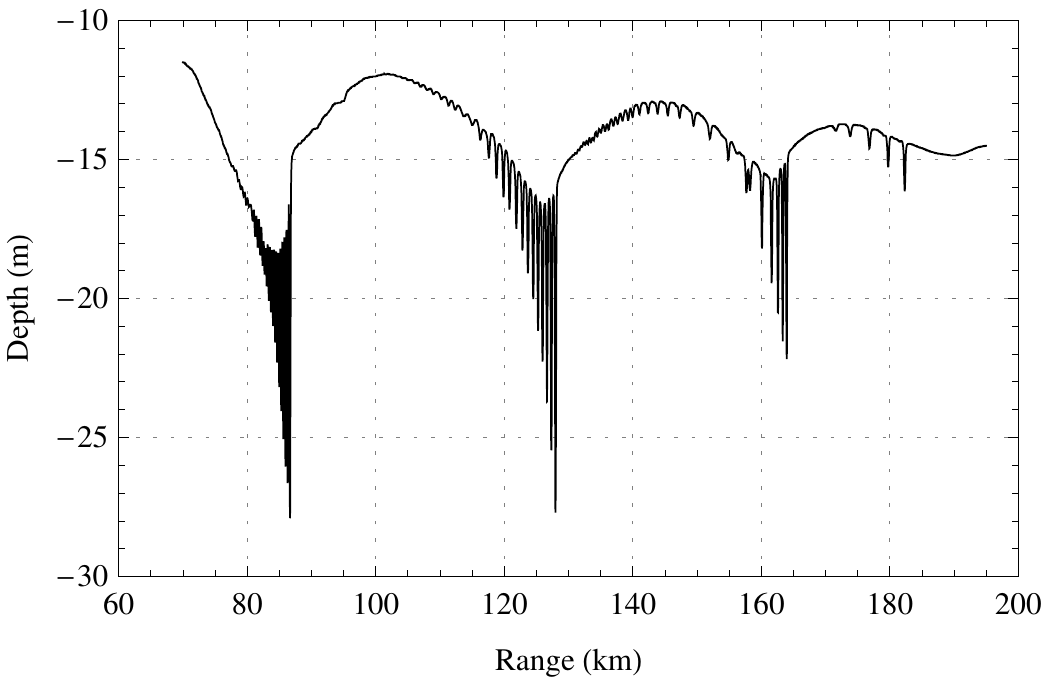}\hfill\includegraphics*[width=0.48\textwidth]{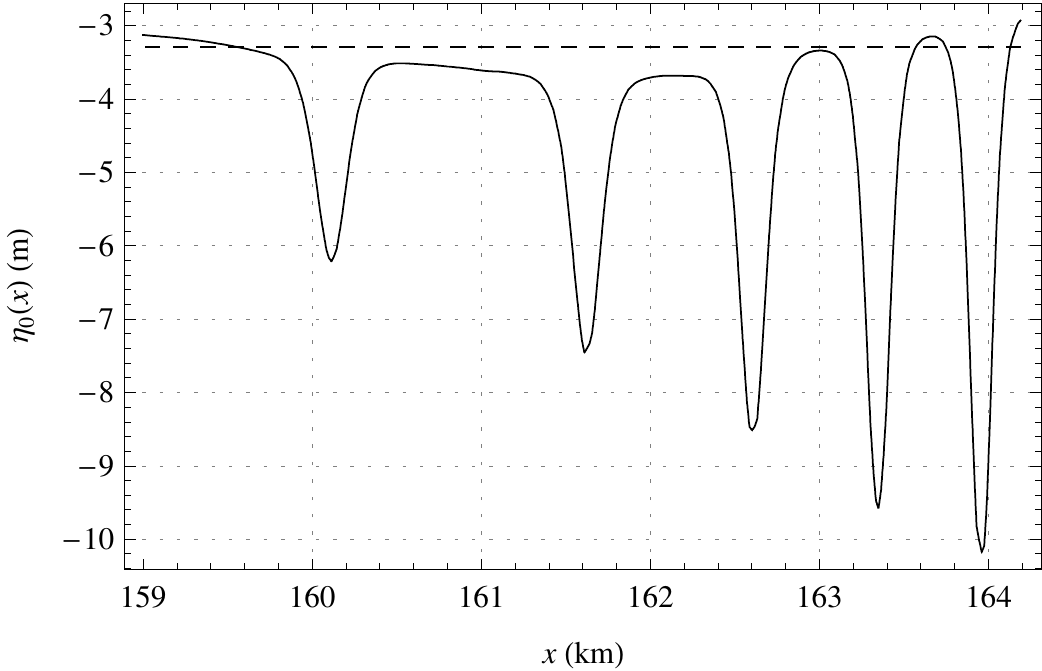}}
\caption{In the left panel, solitary wave packets, generated by the nonhydrostatic Lamb model \cite{l94}, are shown traveling on the $\sigma_t=22.0$ isopycnal in the (simulated) Yellow Sea. In the right panel, the fourth wave packet from the left (i.e., the range from $159.0$ to $164.2$ km) is displayed; the zero depth in the graph corresponds to a total depth of $-h_2(=-12\;\mathrm{m})$ on the left.}
\label{fig:ist_ic}
\end{figure}

In this section, we use the PIST, as described above, to carry out an analysis, along the lines of those in \cite{osbc91,op94,zh99}, of a shallow-water internal solitary wave train from the Yellow Sea simulation studies of Chin-Bing et al.~\cite{cb03} and Warn-Varnas et al.~\cite{wv05}. The complete data set and a zoom of the wave packet that we study [i.e., the initial condition $\eta_0(x)$ for the KdV equation] are shown in Fig.~\ref{fig:ist_ic}. Furthermore, the physical constants needed to compute the parameters in Eq.~\eqref{eq:kdv_params} are taken to be
\begin{equation}
g=9.81\;\mathrm{m/s}^2,\quad h_1=12\;\mathrm{m},\quad h_2=58\;\mathrm{m},\quad \rho_1=1020\;\mathrm{kg/m}^3,\quad \rho_2=1024.9\;\mathrm{kg/m}^3,
\end{equation}
in accordance with those used in \cite{cb03,wv05}. 

These values result in a nonlinearity-to-dispersion ratio $\lambda = -0.00014244$, showing that dispersive effects are significant, as required for the validity of the KdV-based model. In addition, we can compute an Ursell number (see, e.g., \cite{op94}), which is defined as
\begin{equation}
\mathrm{Ur} = \frac{3}{16\pi^2}\left(\frac{\eta_{\mathrm{max}}}{h_2}\right)\left(\frac{L}{h_2}\right)^2,
\label{eq:ursell}
\end{equation}
for the wavetrain to determine its ``degree of nonlinearity.'' For the wave packet shown in the right panel of Fig.~\ref{fig:ist_ic}, we have $\mathrm{Ur}=26.8323$, where $\mathrm{Ur}\simeq1$ is the typical upper limit of linear theory. Therefore, given the latter values of the nonlinearity-to-dispersion ratio and  the Ursell number, it is clear that the wave train under consideration here is quite nonlinear, with dispersive effects dominating. Also, it is interesting to note that the wave packets in Fig.~\ref{fig:ist_ic} (left) are reminiscent of those that emerge from harmonic initial data in the zero-dispersion limit of the KdV equation \cite{gk07}. Indeed, the dimensionless dispersion coefficient is $\propto L^{-3}$, which is a very small quantity for the internal ocean waves.

The final test needed to determine the validity of the Osborne--Burch KdV-based model of internal waves for this data set is to check whether the fundamental assumption of long waves over shallow water, outlined at the end of Section \ref{sec:intsols_model}, is justified. Clearly, the waves are ``long'' as we have $h_1/W\simeq12/200=0.06\ll1$, but the ``shallowness'' assumption is barely satisfied as $[\max_x\eta_0(x)]/h_1\simeq 7/12\approx 0.6$. In fact, it is easy to see that the latter assumption is not valid for the other wave packets featured in the left panel of Fig.~\ref{fig:ist_ic}; they are ``too nonlinear'' to be described by the KdV-based model. This serves to outline the extent to which the nonlinear Fourier analysis is applicable, in general, since it is based on the KdV equation, therefore the evolution of the data should be governed by the latter also.
\begin{figure}
\centerline{\includegraphics*{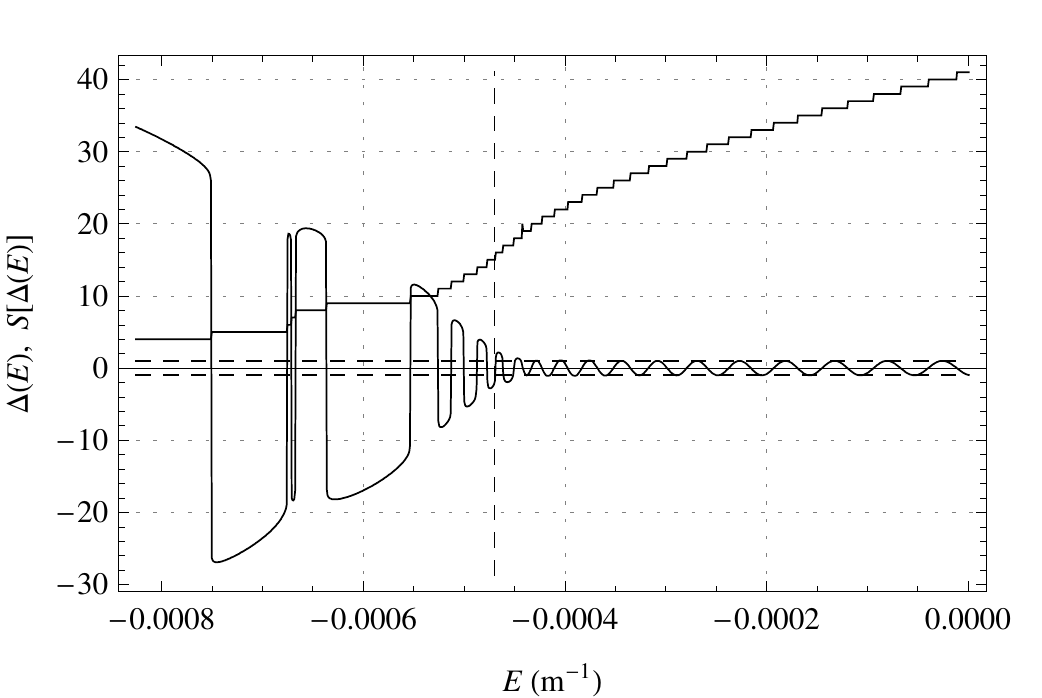}}
\caption{The Floquet diagram, which graphically describes the scattering transform spectrum, of the data in Fig.~\ref{fig:ist_ic}. $\Delta(E)$ is plotted logarithmically outside of the interval $[-1,1]$, which is denoted by the horizontal dashed lines. The staircase solid line is the accounting function of the Floquet discriminant, i.e., $S[\Delta(E)]$.}
\label{fig:ist_fd}
\end{figure}

Nonetheless, even though the wave packet under consideration is barely within the range of validity of the internal solitary waves model described in Section \ref{sec:intsols_model}, we obtain very good results using the scattering transform. To this end, in Fig.~\ref{fig:ist_fd}, the Floquet discriminant $\Delta(E)$ is shown. The latter's crossings of the $\pm 1$ horizontal lines determine the main spectrum eigenvalues $\{\mathcal{E}_j\}_{j=1}^{2N+1}$. The dashed vertical line is the spectral reference level $E_{\text{ref}}$, which delineates the solitons part of the spectrum from the radiation (and other ``less nonlinear'' waves) part. Using the spectral reference level, we compute the reference depth $\eta_{\text{ref}}=-E_{\text{ref}}/\lambda=-3.29767\;\mathrm{m}$ upon which the solitons propagate. The latter is represented in the right panel of Fig.~\ref{fig:ist_ic} by the dashed horizontal line. Physically, the reference depth of the solitons can be understood as follows: in the absence of radiation, $\eta_{\text{ref}}$ would be the depth at which the unperturbed isopycnal, i.e., the the interface between the two layers (recall Fig.~\ref{fig:int_sols}), is located.

\begin{figure}
\centerline{\includegraphics*[width=0.48\textwidth]{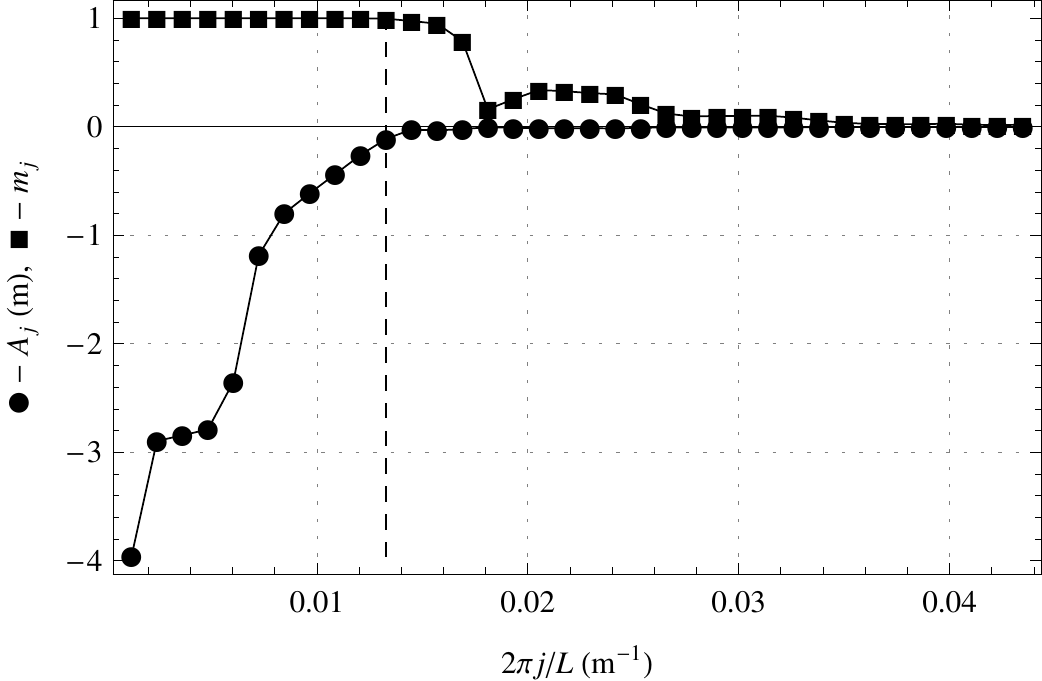}\hfill\includegraphics*[width=0.48\textwidth]{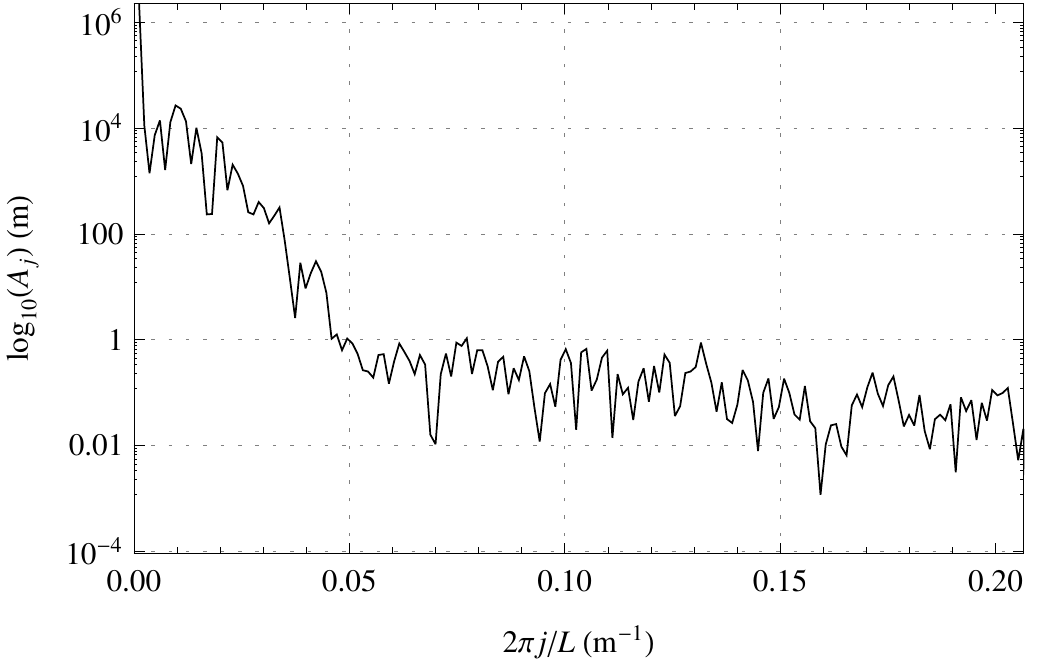}}
\caption{In the left panel is the the PIST spectrum of the data in the right panel of Fig.~\ref{fig:ist_ic}. Squares represent the moduli $m_j$ and circles represent the amplitudes $A_j$ of the hyperelliptic oscillation modes. In the right panel is the (ordinary) Fourier transform spectrum of the same data set. Note that the vertical scale is logarithmic.}
\label{fig:spectrum}
\end{figure}

In Fig.~\ref{fig:spectrum} (left) the amplitudes and moduli of the nonlinear oscillations modes are plotted versus the (commensurable) wavenumbers $k_j=2\pi j/L$. For the data set considered here, $L=5203.88\;\mathrm{m}$, $M=342$ and $N=36$. The dashed vertical line corresponds to the wavenumber $k_{\text{ref}}=2\pi N_{\text{sol}}/L=0.0132814\;\mathrm{m}^{-1}$, where $N_{\mathrm{sol}}=11$ is the number of solitons. It delineates the soliton part of the spectrum from the radiation part, just like the vertical dashed line in Fig.~\ref{fig:ist_fd}. It is clear that the spectrum corresponding to the data in the right panel of Fig.~\ref{fig:ist_ic} consists of eleven solitons. Moreover, there is an assortment of moderately nonlinear waves and radiation; however, the amplitudes of the latter modes are very small. Therefore, the solitons are the dominant part of the spectrum, and the remaining oscillations do not contribute significantly to the physics at hand.

Another important finding is that there are ``hidden'' solitons in the data set. That is to say, in Fig.~\ref{fig:ist_ic} (right) there are clearly four soliton-looking waves, however, the scattering transform spectrum, presented in Fig.~\ref{fig:spectrum} (left), shows there are eleven solitons present. Though the first four are clearly the largest ones, and therefore perceptible to the naked eye, there are seven other ones with much smaller amplitudes, and likely larger ``wavelengths'' (widths), that are indiscernible in the data set. This kind of information, which cannot be obtained by means of the (ordinary) Fourier transform, is important when studying the effects of internal solitary waves on underwater acoustic signals \cite{wv03,zzr91}.

Also in Fig.~\ref{fig:spectrum} (right), we have plotted the (ordinary) Fourier spectrum of the data set under consideration. Note that we cannot easily read off from the plot the number and type of waves that evolve out of the data, like we did above with the scattering transform spectrum. Furthermore, the (ordinary) Fourier spectrum suggest that the most energetic modes are in the range, roughly, $k\in[0,0.05]$. Meanwhile, the scattering transform spectrum shows that the most energetic modes are in the range $k\in[0,k_{\mathrm{ref}}]\approx[0,0.014]$, which is \emph{a fourth} of the range predicted by the (ordinary) Fourier transfrom. Given the nonlinear nature of the phenomenon the data set describes, we can expect the latter wavenumber range to be much more accurate than the former. This ability of the PIST to ``pinpoint'' the solitons' wavenumbers is crucial in studying the acoustic energy transfer (mode conversion) caused by internal waves in the ocean \cite{wv03,zzr91}.

At this point, it is important to note that the simulated data analyzed in this paper is ``clean,'' in the sense that any experimental noise and competing physical effects, other than the generation of solitary solitary waves by the tidal flow over bottom topography, are absent. Consequently, this analysis serves as a benchmark for the application of the PIST and Osborne's nonlinear Fourier analysis to internal solitary wave data. Now that the limits of applicability have been outlined, and the various computational challenges identified and resolved, the PIST is ready for use in the study of \emph{experimental} data, such as the one collected in Strait of Messina and studied by Warn-Varnas et al.~\cite{wv03}.

\section{Perspective and outlook}
\label{sec:perspectives}

Finally, we mention some possible avenues of future work. An important step towards the wide use of the PIST in the oceanographic community is the critical evaluation and benchmarking of the latter against the standard data analysis techniques currently in use, i.e., the windowed Fourier and the continuous wavelet transforms. Such work is underway by Hawkins et al.~\cite{hwvc07}. Another interesting problem is the development of a ``windowed'' PIST. That is to say, for internal solitary wave trains (such as the one shown in the left panel for Fig.~\ref{fig:ist_ic}, for example) we can compute the scattering transform spectrum of \emph{each} wave packet. Thus, we would have a \emph{sequence} of spectra illustrating the ``aging'' process the solitons undergo. Much in the same way the windowed Fourier transforms allows for the simultaneous time and frequency resolution of features, a windowed PIST would allow for the identification of constants of the motion, i.e., the ``nature'' of the soliton evolution. The first step in this direction was taken by Zimmerman and Haarlemmer \cite{zh99}, who used this type of ``stopwatch'' approach to obtain the leading-order invariant of motion of their data. Furthermore, such an approach would enable the study of the process by which solitons are created by tidal flow over topography, along the lines of the work of Osborne et al.~\cite{oosb98}. 

Another issue that deserves further study is the numerical algorithm for finding the eigenvalues of Eq.~\eqref{eq:schro}. The method employed herein, like the original automatic algorithm of Osborne \cite{o94}, relies on the fact that the data set (i.e., the potential of the Schr\"odinger equation) is piecewise constant to obtain Eq.~\eqref{eq:psi_pwc_potent}. This, of course, results in a formally first-order method for the spectral eigenvalues. In this regard, there is a promising new approach due to Deconinck et al.~\cite{dkck07}. The idea is to use a Fourier spectral method in conjunction with Floquet theory to compute the eigenvalues (and eigenfunction) of (linear) differential operators such as $\mathcal{H}$ in Eq.~\eqref{eq:schro}. This approach offers greater flexibility and improved accuracy, therefore its utility in the nonlinear Fourier analysis of physical wave data should be established. 

\bigskip{\bf Acknowledgments}\medskip

First, the author would like to express his gratitude to the referees, whose incisive comments helped improve the manuscript greatly. Second, this work was supported by the U.S.~Naval Research Laboratory (PE 061153N) and the American Society for Engineering Education through the Student Summer Employment Program (Summer 2005) and the Naval Research Enterprise Intern Program (Summer 2007). Third, a Student Award from Prof.~Thiab Taha, organizer of the Fifth IMACS International Conference on Nonlinear Evolution Equations and Wave Phenomena: Computation and Theory, which allowed the author to present this work at the latter conference, is kindly acknowledged. Last but not least, the author would also like to thank Drs. Stanley Chin-Bing, Alex Warn-Varnas and Pedro Jordan for introducing him to this problem, numerous insightful discussions and their encouragement. 

\bibliographystyle{elsart-num-sort}
\bibliography{internal_solitons}

\begin{thebibliography}{10}
\expandafter\ifx\csname url\endcsname\relax
  \def\url#1{\texttt{#1}}\fi
\expandafter\ifx\csname urlprefix\endcsname\relax\def\urlprefix{URL }\fi

\bibitem{as81}
M.~J. Ablowitz, H.~Segur, Solitons and the Inverse Scattering Transform, SIAM,
  Philadelphia, 1981.

\bibitem{a03}
J.~R. Apel, A new analytical model for internal solitons in the ocean, J. Phys.
  Oceanogr. 33 (2003) 2247--2269.

\bibitem{b82}
J.~P. Boyd, Theta functions, {Gaussian} series, and spatially periodic
  solutions of the {Korteweg--de Vries} equation, J. Math. Phys. 23 (1982)
  375--387.

\bibitem{bh91}
J.~P. Boyd, S.~E. Haupt, Polycnoidal waves: {Spatially} periodic
  generalizations of multiple solitons, in: A.~R. Osborne (ed.), Nonlinear
  Topics in Ocean Physics, Elsevier, Amsterdam, 1991, pp. 827--856.

\bibitem{c99}
C.~Chicone, Ordinary Differential Equations with Applications, vol.~34 of Texts
  in Applied Mathematics, Springer--Verlag, New York, 1999.

\bibitem{cb03}
S.~A. Chin-Bing, A.~Warn-Varnas, D.~B. King, K.~G. Lamb, M.~Teixeira, J.~A.
  Hawkins, Analysis of coupled oceanographic and acoustic soliton simulations
  in the {Yellow Sea: A} search of soliton-induced resonances, Math. Comput.
  Simul. 62 (2003) 11--20.

\bibitem{db74}
G.~Dahlquist, \r{A}. Bj\"orck, Numerical Methods, Dover Publications, New York,
  2003.

\bibitem{dkck07}
B.~Deconinck, F.~Kiyak, J.~D. Carter, J.~N. Kutz, {SpectrUW: A} laboratory for
  the numerical exploration of spectra of linear operators, Math. Comput.
  Simul. 74 (2007) 370--378.

\bibitem{gk07}
T.~Grava, C.~Klein, Numerical solution of the small dispersion limit of
  {Korteweg--de Vries} and {Whitham} equations, Comm. Pure Appl. Math. 60
  (2007) 1623--1664.

\bibitem{hwvc07}
J.~A. Hawkins, A.~Warn-Varnas, I.~Christov, {Fourier}, scattering, and wavelet
  transforms: {Applications} to internal gravity waves with comparisons to
  linear tidal data, in: R.~V. Donner, S.~M. Barbosa (eds.), Nonlinear Time
  Series Analysis in the Geosciences: {Applications} in Climatology,
  Geodynamics and Solar-Terrestrial Physics, vol. 112 of Lecture Notes in Earth
  Sciences, Springer, Berlin/Heidelberg, 2008, pp. 223--244.

\bibitem{l94}
K.~G. Lamb, Numerical experiments of internal wave generation by strong tidal
  flow across a finite amplitude bank edge, J. Geophys. Res. 99~(C1) (1994)
  843--864.

\bibitem{oo91}
A.~R. Osborne, Nonlinear {Fourier} analysis, in: A.~R. Osborne (ed.), Nonlinear
  Topics in Ocean Physics, Elsevier, Amsterdam, 1991, pp. 669--699.

\bibitem{o91}
A.~R. Osborne, Nonlinear {Fourier} analysis for the infinite-interval
  {Korteweg--de Vries} equation {I: An} algorithm for the direct scattering
  transform, J. Comput. Phys. 94 (1991) 284--313.

\bibitem{o93}
A.~R. Osborne, Numerical construction of nonlinear wave-train solutions of the
  periodic {Korteweg--de Vries} equation, Phys. Rev. E 48 (1993) 296--309.

\bibitem{o94}
A.~R. Osborne, Automatic algorithm for the numerical inverse scattering
  transform of the {Korteweg--de Vries} equation, Math. Comput. Simul. 37
  (1994) 431--450.

\bibitem{o95}
A.~R. Osborne, Solitons in the periodic {Korteweg--de Vries} equation, the
  {$\Theta$}-function representation, and the analysis of nonlinear, stochastic
  wave trains, Phys. Rev. E 52 (1995) 1105--1122.

\bibitem{ob86}
A.~R. Osborne, L.~Bergamasco, The solitons of {Zabusky} and {Kruskal}
  revisited: {Perspective} in terms of the periodic spectral transform, Physica
  D 18 (1986) 26--46.

\bibitem{ob80}
A.~R. Osborne, T.~L. Burch, Internal solitons in the {Andaman Sea}, Science 208
  (1980) 451--460.

\bibitem{oosb98}
A.~R. Osborne, M.~Onorato, M.~Serio, L.~Bergamasco, Soliton creation and
  destruction, resonant interactions, and inelastic collisions in shallow water
  waves, Phys. Rev. Lett. 81 (1998) 3559--3562.

\bibitem{op94}
A.~R. Osborne, M.~Petti, Laboratory-generated, shallow-water surface waves:
  {Analysis} using the periodic, inverse scattering transform, Phys. Fluids 6
  (1994) 1727--1744.

\bibitem{os90}
A.~R. Osborne, E.~Segre, Numerical solutions of the {Korteweg--de Vries}
  equation using the periodic scattering transform {$\mu$}-representation,
  Physica D 44 (1990) 575--604.

\bibitem{osbc91}
A.~R. Osborne, E.~Segre, G.~Boffetta, L.~Cavaleri, Soliton basis states in
  shallow-water ocean surface waves, Phys. Rev. Lett. 67 (1991) 592--595.

\bibitem{sep03}
A.~Salupere, J.~Engelbrecht, P.~Peterson, On the long-time behaviour of soliton
  ensembles, Math. Comput. Simul. 62 (2003) 137--147.

\bibitem{wv03}
A.~C. Warn-Varnas, S.~A. Chin-Bing, D.~B. King, Z.~Hallock, J.~A. Hawkins,
  Ocean-acoustic solitary wave studies and predictions, Surv. Geophys. 24
  (2003) 39--79.

\bibitem{wv05}
A.~C. Warn-Varnas, S.~A. Chin-Bing, D.~B. King, J.~A. Hawkins, K.~G. Lamb,
  M.~Teixeira, {Yellow Sea} ocean-acoustic solitary wave modeling studies, J.
  Geophys. Res. 110 (2005) C08001.

\bibitem{zk65}
N.~J. Zabusky, M.~D. Kruskal, Interaction of ``solitons'' in a collisionless
  plasma and the recurrence of initial states, Phys. Rev. Lett. 15 (1965)
  240--243.

\bibitem{zzr91}
J.-X. Zhou, X.-Z. Zhang, P.~H. Rogers, Resonant interaction of sound wave with
  internal solitons in the coastal zone, J. Acoust. Soc. Am. 90 (1991)
  2042--2054.

\bibitem{zh99}
W.~B. Zimmerman, G.~W. Haarlemmer, Internal gravity waves: {Analysis} using the
  the periodic, inverse scattering transform, Nonlinear Process. Geophys. 6
  (1999) 11--26.

\end{thebibliography}
\end{document}